\documentclass[draft,grl]{agu2001}

\usepackage [xdvi]{graphicx}

\authorrunninghead{RICCI, LAPENTA, AND BRACKBILL}
\titlerunninghead{Magnetotail plasma sheet structure}

\journalid{????} \articleid{???}{??} \paperid{????}
\cpright{AGU}{2001}

\received{????} \revised{????}
\accepted{????} \published{}

\authoraddr{P. Ricci,
Dipartimento di Energetica, Politecnico di Torino, Corso Duca
degli Abruzzi 24 - 10129 Torino, Italy.}

\authoraddr{G. Lapenta, Los Alamos National Laboratory, Los Alamos NM 85744
(lapenta@lanl.gov).}

\authoraddr{J.U. Brackbill, Department of Mathematics and Statistics,
University of New Mexico, Albuquerque NM.}

\begin{document}

\title{Plasma sheet structure in the magnetotail:
kinetic simulation and comparison with satellite observations}

\author{Paolo Ricci$^{(1,2)}$, Giovanni Lapenta$^{(1,3)}$ and J. U. Brackbill$^{(4)}$}

\affil{
(1) Istituto Nazionale per la Fisica della Materia (INFM), Dipartimento di Fisica,
Politecnico di Torino, Torino, Italy \\
(2) Dipartimento di Energetica, Politecnico di Torino, Torino, Italy\\
(3) Theoretical Division, Los Alamos National Laboratory, Los
Alamos NM, USA\\
(4) Department of Mathematics and Statistics, University of New
Mexico, Albuquerque NM}

\begin{abstract}

We use the results of a three-dimensional kinetic simulation of an
Harris current sheet to shown and reproduce the ISEE-1/2, Geotail,
and Cluster observations of the magnetotail current sheet
structure. Current sheet flapping, current density bifurcation,
and reconnection are explained as the results of the
self-consistent evolution of a Harris current sheet, where
lower-hybrid drift, kink, and tearing instabilities are involved.

\end{abstract}

\begin{article}

\section{Introduction}

The magnetotail current sheet is one of the key topics in
magnetospheric physics.

Observations of the current sheet have revealed a very complex
structure. At the end of April 2, 1978, the ISEE-1/2 spacecraft
detected a flapping of the plasma sheet and the spacecraft crossed
the central region more than 10 times in a hour. In particular,
during a "turbulent" crossing, the spacecraft detected current
concentration outside the central region, unlike the Harris
current sheet [{\it Sergeev et al.}, 1993]. In fact, Geotail [{\it
Kokobun et al.}, 1994; {\it Mukai et al.}, 1994] averaged data
obtained from October 1993 to June 1995 show that the structure of
the plasma sheet can be often approximated by a double-peaked
electric current sheet [{\it Hoshino et al.}, 1996] and
observations made by the same spacecraft during a substorm on 23
April 1996 lead to a similar conclusion [{\it Asano et al.}, 2003].
On January 14, 1994, Geotail also detected multiple double-peaked
current sheet crossings, associated with plasma flow [{\it Hoshino
et al.}, 1996]. More recently, time analysis of data from the four
Cluster spacecrafts [{\it Balogh et al.}, 2001] showed that fast
motion and bifurcation of the current sheet are associated with a
wave-like transient propagating in the dawn-to-dusk direction
[{\it Sergeev et al.}, 2003; {\it Runov et al.}, 2003]. Plasma
flow has also been observed during a substorm event [{\it Hoshino et
al.}, 1996; {\it \O ieroset et al.}, 2001; {\it Asano et al.},
2003]. These observations refer both to the distant magnetotail
($\approx 100 R_E$) [{\it Hoshino et al.}, 1996] and to a region
closer to Earth ($\approx 15 R_E$) [{\it Sergeev et al.}, 1993;
{\it Asano et al.}, 2003; {\it Runov et al.}, 2003; {\it Sergeev
et al.}, 2003]

A useful but simple one-dimensional description of the current
sheet is given by the Harris model, where the magnetic field is
given by $B_x(z)=B_0 \tanh(z / \lambda)$ and the plasma
density, proportional to the current
density, is given by $n(z)=n_0 \cosh^{-2} (z/ \lambda)$, where
$\lambda$ is the half thickness of the current sheet and the GSM
coordinates are used.

Generalizations of the standard Harris current sheet equilibrium
have been recently proposed to reproduce the bifurcation observed
by satellites [{\it Shindler and Birn}, 2002; {\it Sitnov et al.},
2003]. Runov {\it et al.} [2003] propose that bifurcation and
flapping signatures are due to a kink or a sausage instability of
a current sheet with an enhanced current density on both
hemispheres. Zelenyi {\it et al.} [2002] show that non-adiabatic
effects can reduce the current density in the center of the
current sheet. A bifurcated current sheet can be present in the
plasma outflow region when magnetic reconnection is occurring
[e.g., {\it Arzner and Sholer}, 2001]. Karimabadi {\it et al.}
[2003a, 2003b] argue that the ion-ion kink instability causes a
displacement of the current sheet which can explain the flapping
observations and interpret the bifurcated structure of the current
sheet as a magnetic field profile with weak central gradient due
to the non-linear evolution of the kink instability. Generally,
plasma flow is explained in terms of plasma out-flowing from a
reconnection region.

The aim of the present work is to analyze the results of a
three-dimensional kinetic simulation of the Harris current sheet
by introducing diagnostic tools very similar to the one used by
satellite. We show that the self-consistent evolution of the
current sheet dispalyed by the simulations can be responsible
for the data observed by the satellites described in the
references above. In particular, taking into account the relative
motion of the current sheet and the spacecraft, and the Cluster
tetrahedron configuration, we recover the most significant
magnetic data records obtained by the Cluster spacecraft as the
signature of current sheet flapping. The recurrence frequency of
the magnetic field $B_x$ allow the comparison with observations by
GEOTAIL [{\it Hoshino et al.}, 1996] that show current
bifurcation, and the magnetic field structure show the signature
of bifurcation recovered in single crossing. We also compare the
plasma flow due to the tearing instability with the observations.

\section{The simulations}

In our study, we use the implicit PIC code CELESTE3D [{\it
Forslund and Brackbill}, 1985; {\it Vu and Brackbill}, 1992; {\it
Ricci et al.}, 2002a], which is particularly suitable for large
scale and long period kinetic simulations performed with high mass
ratio and has been applied previously to problems in
magnetospheric physics [e.g., {\it Lapenta and Brackbill}, 2000;
{\it Lapenta and Brackbill}, 2002; {\it Ricci et al.}, 2002b; {\it
Lapenta et al.}, 2003].

We use the same plasma parameters as the GEM challenge [{\it Birn
et al.}, 2001]. In particular we start from a standard Harris
current sheet, with magnetic field given by $B_x(z)=B_0 \tanh (z/
\lambda)$, density by $n(z)=n_0\cosh^{-2} (z/ \lambda)+ n_b$ with
$\lambda=0.5 c/\omega_{pi}$, $T_i/T_e=5$, the ion drift velocity
is $V_{i0}=1.67 V_A$, and a background population $n_b=0.2 n_0$.
Unlike the GEM challenge, we do not add any initial perturbation
and let the system evolve on its own. The dimensions of the system
are $[-L_x/2, L_x/2] \times [-L_y/2, L_y/2] \times [-L_z/2,
L_z/2]$ with $L_x= 12.8 c / \omega_{pi}$, $L_y = 19.2 c /
\omega_{pi}$, and $L_z = 6.4 c / \omega_{pi}$, discretized with a
grid $N_x \times N_y \times N_z =32 \times 48 \times 32$. The
boundary conditions are perfect conductors at $z=\pm L_z$ and
periodic boundaries in all the other directions. The mass ratio is
$m_i/m_e=180$. The parameters chosen make the current sheet
particularly unstable and its dynamics are accelerated compared
with typical magnetotail current sheets. We are constrained to do
that in order to follow the dynamics of the current sheet in a
reasonable computational time. As a consequence, it is necessary
to scale our results to make a quantitative comparison between
simulation results and observations. In any case, the general
trends can be located and the linear theory [{\it Karimabadi et
al.}, 2003a] can help in scaling the results.

Previous simulations [{\it Lapenta and Brackbill}, 2002; {\it
Lapenta et al.}, 2003; {\it Daughton}, 2003] performed in the
current aligned plane show, in absence of a plasma background, the
development of the fastest lower-hybrid drift instability on the
electron gyroscale, followed by electromagnetic modes with
wavelengths intermediate between the ion and the electron
gyroscale. The lower hybrid drift instability causes a velocity
shear (present since the beginning of the simulation when a
background plasma is present) that triggers a Kelvin-Helmhotz (KH)
instability that kinks the current sheet. As we add a background
population, following Karimabadi {\it et al.} [2003b], the
velocity shear is present since the beginning of the simulation
and the resulting KH instability can be also interpreted as a
kinetic ion-ion kink instability [{\it Karimabadi et al.}, 2003a;
{\it Karimabadi et al.}, 2003b].

Both the present and previous simulations have shown that with the
kinking of the current sheet a tearing instability grows [{\it
Lapenta and Brackbill}, 2001; {\it Lapenta et al.}, 2003]. In the
present case, the fastest short wavelength modes grow
(corresponding to $k_x L \approx 0.5$ or $m_x=2$ in our simulation
box), and then they merge to form a single island tearing mode
that involve the whole domain.

Below we consider specific aspects of the satellite observations
basing their interpretation on the simulation.

\section{Current sheet flapping}

When vertical oscillations of the plasma sheet (flapping) occur,
spacecrafts may repeatedly cross the current sheet. Clear evidence
of current sheet flapping is shown by ISEE-1/2 [{\it Sergeev et
al.}, 1996], by Geotail [{\it Hoshino et al.}, 1996], and by
Cluster [{\it Runov et al.}, 2003; {\it Sergeev et al.}, 2003]. We
focus our attention on the observations by Cluster and show that
current sheet kinking developed in the course of our simulations
can explain those.

Figure 1 shows fully developed current sheet kinking. The $B_x$
field is shown. The wavelength is $k_y \lambda \approx 0.5$, which
matches fairly well the observed wavelength in Runov {\it et al.}
[2003] ($k_y \lambda =0.7$). The linear theory predicts a
decrease of the wavelength when $\rho_i/\lambda$ increases [{\it
Karimabadi et al.}, 2003a] consistent with the fact that our thickness
is likely smaller than the observation. The amplitude $A$ at time
$t\omega_{ci}=16$ is $A/\lambda \approx 2$ is comparable to the
observed value ($A/\lambda \approx 1.4$) [{\it Sergeev et al.},
2003]. The flapping motion observed by Cluster moving duskward at
$v_{ph} \approx 200$ km/s, corresponding to approximatively $0.2
v_A$. The kink instability shown in our simulations gives a
$v_{ph,SIM} \approx 0.5 v_A$, larger than observed in space.
However, the linear theory predicts a decrease of the phase
velocity when $\rho_i/\lambda$ increases. Since we use an
artificially high $\rho_i/\lambda$, the higher phase speed is
justified and consistent with our interpretation of the flapping
motion.

In Fig. 2a we show Cluster $\#2$ and $\#3$ observations taken on
29 August 2001, which have been analyzed previously by Runov {\it
et al.} [2003]. In particular, the $B_x$ data is considered. In
Fig. 3a, we evaluate the magnetic field as a function of time as
would be recorded by a virtual spacecraft placed in the
environment provided by the simulation. According to the real
spacecraft disposition, we impose a distance between the two
virtual satellites in the $z$ direction to be of the order of
$\lambda/2$. Cluster observes an oscillation period of $\tau=90$s
and a relative velocity between satellite and plasma $v_{ph}
\approx 0.2 v_A$. In order to decrease the time necessary for the
observation, we increase the relative satellite velocity up to
$v_{SIM} = 5 v_A$, thus decreasing the oscillation period to
$\tau_{SIM} = 2 \omega_{ci}^{-1}$, in good agreement with the
oscillation period recorded by Cluster. With the new relative
velocity and using the fact that $\omega_{ci} \approx 0.6 s^{-1}$
in the magnetotail and the observed period is of the order $54
\omega_{ci}^{-1}$, the observed wavelength, $v_{ph} \tau \approx
11 c/\omega_{pi}$ and the simulated wavelength, $v_{SIM}
\tau_{SIM} \approx 10 c/\omega_{pi}$, are comparable.

The flapping observation recorded by Cluster $\#3$ on September
26, 2001 and described by Sergeev {\it et al.} [2003] is shown in
Fig. 2b. It is reproduced by our simulations at later times, when
the amplitude of the kink has grown enough that the virtual
satellite can pass from one side to the other of the current
sheet. This is shown in Fig. 3b. We note that Cluster observations
reveal a flattening of the current sheet in the vicinity of the
points where $B_x=0$, which is associated to current sheet
bifurcation. The grid spacing used in the three-dimensional
simulation does not allow to resolve this structure, which is
better addressed in more resolved two-dimensional simulations.

In agreement with Sergeev {\it et al.} [2003] and Runov {\it et
al.} [2003], our simulations reveal that the current sheet
flapping is mostly in the $(y,z)$ plane, while the tilt in the
$(x,z)$ plane is insignificant.

\section{Current sheet bifurcation}

Experimental evidence exists that the current distribution in the
magnetotail sheet may be double-peaked, with a pair of electric
current sheets separated by a layer of a weak quasi-uniform
magnetic field. Current sheet bifurcation has been revealed both
in averages over a number of current sheet crossings, and in
single sheet crossings.

The statistical studies of the current sheet presented in Hoshino
{\it et al.} [1996] reveal a bifurcated current profile. An
ensemble of neutral sheet crossings is considered and the
occurrence frequency of $B_x$ is evaluated. The observed
distribution has a peak around the null magnetic region, as in
Sergeev {\it et al.} [2003]. The distribution of occurrence of
the field $B_x$, $N(B_x)$ can be expressed as

\begin{equation}
N(B_x) \propto \frac{d}{d B_x}F^{-1}(B_x)
\end{equation}

where $B_x(z)=F(z)$. Thus, the functional form of the magnetic
field as a function of $z$ can be obtained, and from the gradient
of $B_x$ with respect to $z$ it is possible to evaluate the plasma
current. The whole procedure is able to smear out current sheet
flapping and particular motion of the current sheet.

In order to study current bifurcation, we have performed a
two-dimensional simulation in the $(y,z)$ plane that allows to use
a more refined grid in this plane ($N_y \times N_z= 128 \times
64$). In Fig. 4a we show the plot of in-plane current
$\sqrt{J_y^2+J_z^2}$ at $t\omega_{ci}=20$. Although the high
fluctuations, the simulation show an increase of the current on
the flanks of the current sheet. To follow GEOTAIL data analysis,
we compute the recurrence frequency of $B_x$ (Fig 4b). The
occurrence frequency shows a peak around $B_x=0$, as satellite
data show [{\it Hoshino et al.}, 1996 (Fig.2), {\it Sergeev et
al.}, 2003 (Fig. 4)]. By integrating Eq. (1), we find $B_x(z)$
(Fig. 4c), which is compared with a Harris sheet profile, and the
current profile as a function of $z$ (Fig. 4d). The current is
depleted at the center of the current sheet and two current peaks
grow on the flanks of the initial current sheet. (This is unlike
the Harris sheet equilibrium, where $\partial B_x/\partial z$ is
maximum at $z=0$ where $B_x=0$.) In Fig. 4d the current density
profile is compared with GEOTAIL observations [{\it Hoshino et
al.}, 1996 (Fig.4)] and found in remarkable agreement.

Single crossing observations of current sheet bifurcation are
shown by Runov {\it et al.} [2003] and by Sergeev {\it et al.}
[2003]. We focus on Fig. 3c in Sergeev {\it et al.} [2003], which
shows reduced $\partial B_x / \partial z$ in the central part of
the current sheet (reduced current) and enhanced gradient
(enhanced current density) at the boundary part. In Fig. 5, we
plot a number of $B_x$ profiles as a function of $z$, at different
values of $y$. The same features of the magnetic field structure
shown by satellite observations is recovered.

We finally remark that observations by Geotail on 23 April 1996
show that positive $d | B_x| / dt$ corresponds to relevant current
density $J_y$ [{\it Asano et al.}, 2003]; the same effect is also
recovered within our simulation.

\section{Reconnection}

Not only does a kink instability grow, but also a tearing
instability develops in the Harris sheet, which leads to the
reconnection of the magnetic field lines and outflow and inflow
plasma jets. Satellite observations typically reveal the
reconnection process either by detecting inflow and outflow plasma
jets, which can be very noisy [e.g., {\it Asano et al.}, 2003], or
by detecting earthward and tailward plasma jets with velocities of
the order of $0.1 v_A$ or bigger [{\it Hoshino et al.}, 1996], or
even by detecting flow reversal [{\it \O ieroset et al.}, 2001].
In Fig. 6 we show signatures of magnetic reconnection by showing a
flow reversal associated with a change in the sign of the
reconnecting field. The earthward and tailward velocities,
detected during the crossing of the current sheet, are of the
order of $0.1 v_A$, which roughly corresponds to the typical order
of magnitude of plasma velocity in the satellite observations.

\section{Conclusion}

We have used the results of three-dimensional and two-dimensional
kinetic simulations of Harris current sheet to show that satellite
observations of current sheet flapping, current bifurcation, and
reconnection can all be explained self-consitently. We have
chosen to start from a relatively thin and
unstable current sheet ($\lambda/d_i=0.5$) in order to accelerate
the plasma dynamics. Such thin current sheets are indeed observed
in the magnetotail [e.g., {\it Asano et al.}, 2003, for a review].

Flapping oscillations have been shown as the results a kink
instability that affects the whole current sheet dynamics.
Frequency and amplitude compare well with satellite observations.
Both average and single crossing signatures of current sheet
bifurcation have been detected. Flow reversal, signature of
reconnection, is also shown in presence of a changing sign
$B_z$ component.

\begin{acknowledgments}
The authors gratefully thank M. Hoshino for the permission to use
the data plotted in Fig. 4 and J. Birn, J. Chen, W. Daughton, I.
Furno, M. Taylor, A. Vaivads for helpful discussions. The
satellite data has been obtained from Cluster FGM team [{\it
Balogh et al.}, 2001]. This research is supported by the
Laboratory Directed Research and Development (LDRD) program at the
Los Alamos National Laboratory, by the United States Department of
Energy, under Contract No. W-7405-ENG-36 and by NASA, under the
"Sun Earth Connection Theory Program".
\end{acknowledgments}

\newpage

\begin{itemize}

\item

Fig. 1: The kink of the current sheet is presented by showing the
$x$ component of magnetic field, $B_x$. Both quantities are shown
as a function of $y$ and $z$, at time $t \omega_{ci}= 16$ and at
$x=0$. $B_x$ is normalized to $B_0$.

\item

Fig. 2: Signatures of current sheet flapping, observed by the FGM
Cluster experiment [{\it Balogh et al.}, 2003]. We report the
$B_x$ magnetic field recorded by satellites $\#2$ (dashed) and
$\#3$ (solid) on 29 August 2001 that has been described by Runov
{\it et al.} [2003] (a), and by satellite $\#3$ on September 26
2001, described by Sergeev {\it et al.} [2003] (b).

\item

Fig. 3: Signatures of current sheet flapping as would be recorded
by a virtual spacecraft placed in the environment provided by the
simulation and which reproduce the real signature shown in Fig. 2.
The $B_x$ magnetic field is plotted, normalized to $B_0$.

\item

Fig. 4: Current density  $\sqrt{J_y^2+J_z^2}$ from the
two-dimensional simulation at time $t\omega_{ci}=20$ (a), $B_x$
(normalized to $B_0$) occurrence frequency (b), $B_x$ profile as a
function of $z$ (solid) and comparison with Harris current sheet
(dotted) (the normalization is arbitrary) (c), and current profile
from the simulation compared with Geotail observations [Hoshino
{\it et al.}, 2003, Fig. 4b] (the original dimensionless units
have been scaled to fit the simulation results) (d).

\item

Fig. 5: $B_x$ profile as a function of $z$ for different value of
$y$.

\item

Fig. 6: Typical signature of reconnection: during the crossing of
the current sheet, the reconnecting field, $B_z$, changes sign (a)
and it is associated to earthward and inward plasma jets (b).

\end{itemize}

\end{article}

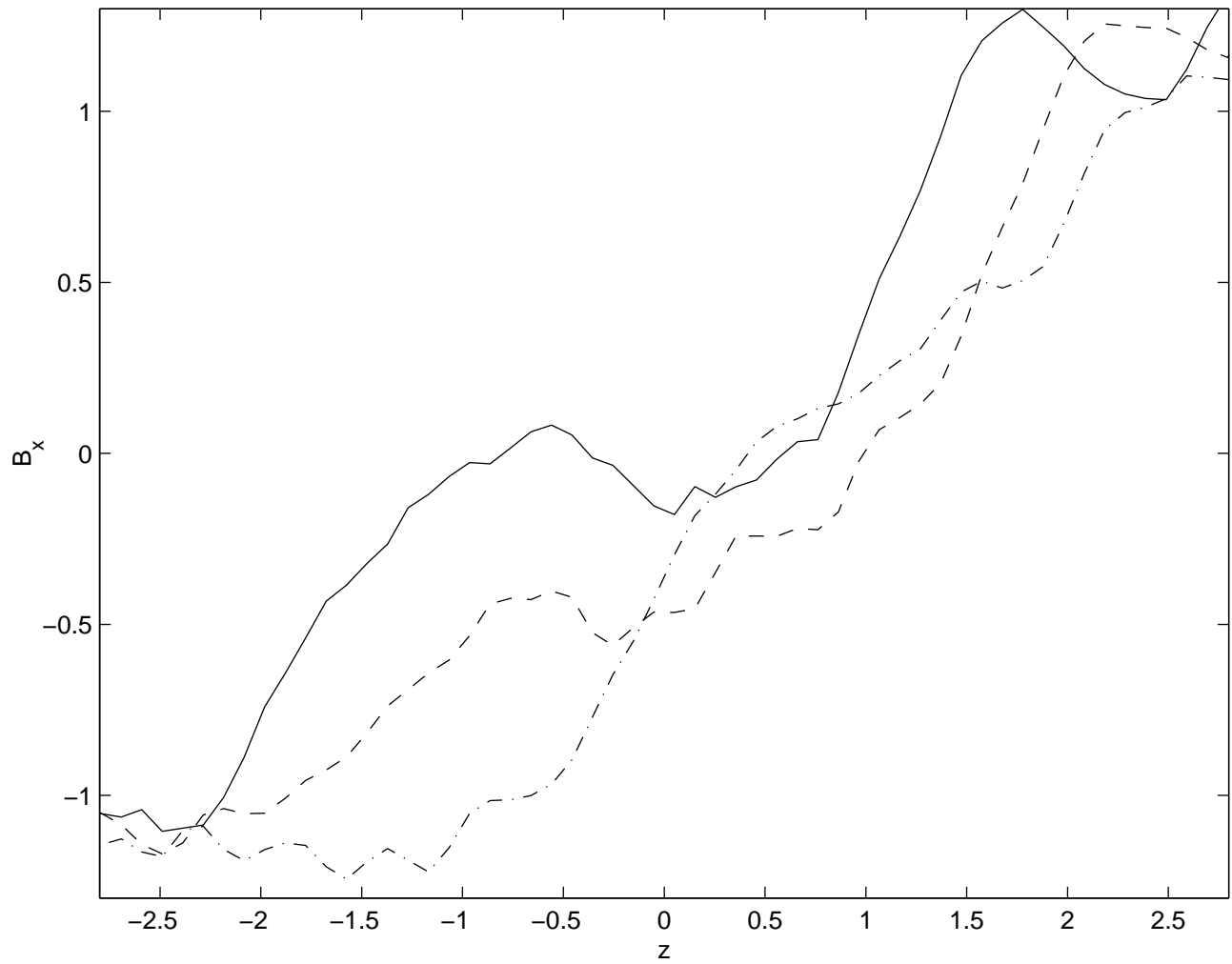
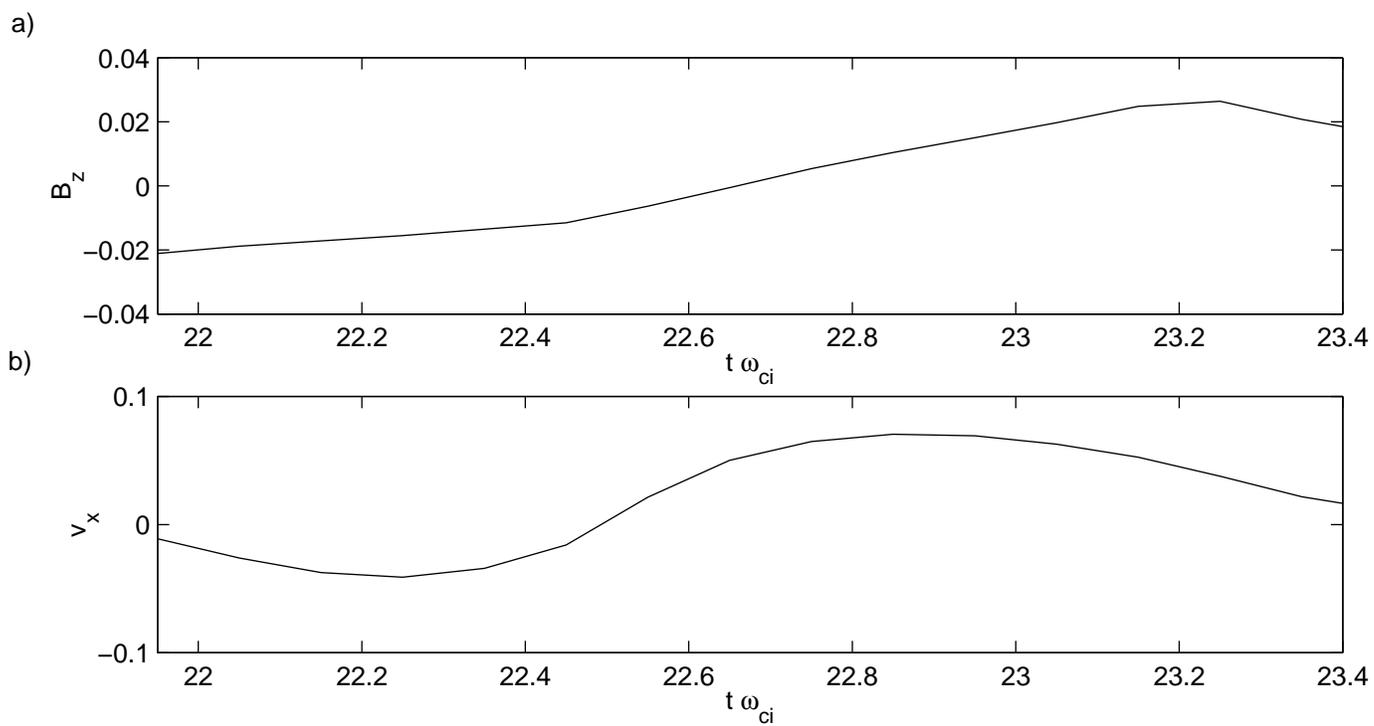
\end{document}